# Distribution fitting 16. How many colors are actually in the field?

**Lorentz JÄNTSCHI**

AcademicDirect, Cluj-Napoca 400641 Cluj, Romania; lori@academicdirect.org

**Abstract.** A study to compare different methods of estimation was conducted. The goal was to provide an estimate for the number of petal colors existing in the field by using a random sample of *Lycoris longituba* flowers taken from the field. The study revealed that the estimation from observed sample agrees very well with third order Jackknife method of estimation.
**Keywords:** Estimating methods; Diversity analysis; Apparent abundance; Natural variability; Petal color

### Introduction

There are many methods estimating the abundance, some of them involving laborious calculations, such as the method of rarefaction. A review of the different approaches could be found in (Gotelli & Colwell, 2001). Unfortunately, most of the studies estimate species richness, but some of them estimate qualitative traits.

The motivation of the study started from the desire to give an estimate for the number of colors of the flowers in the field based on a random sample. Thus, a series of estimating methods were adapted to be used on this concern and were applied on a data series taken from the literature.

The aim of the study was to find one or more alternatives to calculate the number of the colors of the flowers in the field based on a given random sample.

### Material

*Lycoris longituba* is a cultivar of lily (*Amaryllidaceae*) family originated from China. Some authors consider the *Lycoris longituba* a distinct species (Hsu & Fan, 1974; Hsu & others, 1994). Its flowers (see Figure 1) have many colors, being attractive to bees, butterflies and birds.

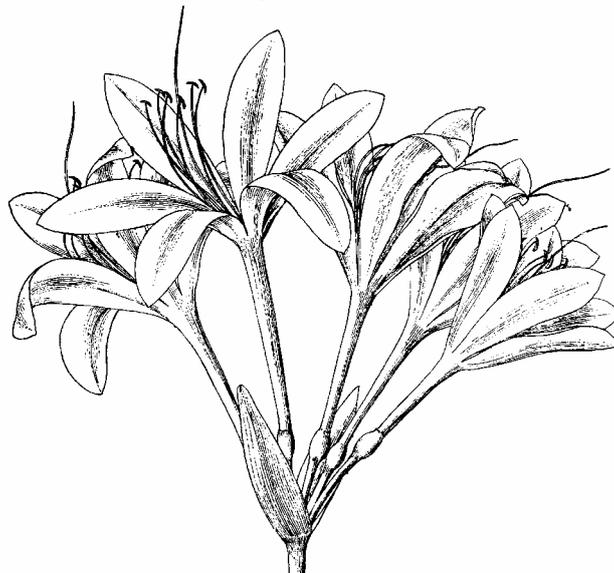

Figure 1. *Lycoris longituba* flower (adapted from Hsu & Fan, 1974)

The data included in this study were taken from (He & others, 2011). Table 1 contains the composition by color of a random sample of 44 flowers sampled randomly from a population in the natural distribution of this species in their natural habitat in China.

Table 1. A random sample of *Lycoris longituba* from the field (according to He & others, 2011)

| Color   | purple | red | yellow | orange | white |
|---------|--------|-----|--------|--------|-------|
| Flowers | 14     | 10  | 10     | 9      | 1     |

**Method**

First method used in this study was derived by Fisher in 1943 (Fisher, 1943) for estimating the abundance of the species. Under assumption of negative binomial distribution of the population subject to sampling, Fisher obtained a limiting form of the distribution excluding zero observations and applicable when a sample drawn from the population did not catch its entire diversity. Let C be the observed number of flower's colors and N the total number of flowers sampled. Then, according to (Fisher, 1943), C and N are in a relationship:

$$C - \alpha \cdot \ln\left(1 + \frac{N}{\alpha}\right) = 0 \qquad (1)$$

with a unique solution ($\alpha = \alpha(C,N)$) with the meaning of richness in colors. Even more, under this assumption of an incomplete sample from negative binomial distribution, the variance of C for given N is:

$$\text{Var}(C) = \alpha \cdot \ln\left(\frac{2N + \alpha}{N + \alpha}\right) - \frac{\alpha^2 N}{(N + \alpha)^2} \qquad (2)$$

Second applied method was the Bootstrap method. The Bootstrap method gives the estimates (Efron, 1979) from a repeated resampling with replacement:

> ÷ Give equal weight (of 1/n for a sample of size n) to each sampled observation;
> ÷ For each *i* from *1* to *N* (usually N from 50 to 200)
>   o Extract (with replacement) a sample of (same) size *n* from the (initial) sample by using discrete uniform distribution and calculate the desired parameter (C) on it (let be $c_i$ its value);
> ÷ The Bootstrap estimates for the parameter (C) and its variance (Var(C)) are:
> $$C_B = \frac{1}{N}\sum_{i=1}^{N} c_i, \quad \text{Var}(C_B) = \frac{1}{N-1}\sum_{i=1}^{N}(C_B - c_i)^2$$

(3)

Third applied method was the Jackknife method. The Jackknife method can be applied by removing systematically one or more observations from the sample (Quenouille, 1956). If only one observation is removed at a time, then *n* sub-samples of *n-1* observations each (then the bias is of O(1/n) order - see Quenouille, 1949) are obtained. If two observations are removed at a time, *n(n-1)/2* sub-samples of *n-2* observations each are obtained. If three observations are removed at a time, *n(n-1)(n-2)/6* sub-samples of *n-3* observations each (and so on) are obtained. For groups up to size *k*, the bias is removed to order $O(1/n^k)$. Following algorithm gives the Jackknife estimates:

> ÷ For any *k* (here from *1* to *3*) do For each *i* from *1* to *Combin(n,k)*
>   o Remove *k* observations from sample and calculate the desired parameter (C) (let be $c_i$ its value);
>   o Compute $CJ_i = (n \cdot C - (n-k) \cdot c_i)/k$;
> ÷ The Jackknife estimates for the parameter and its variance are:
> $$C_{Jk} = \frac{1}{\text{Combin}(n,k)}\sum_{i=1}^{N} CJ_i, \quad \text{Var}(C_{Jk}) = \frac{1}{n \cdot (\text{Combin}(n,k) - k)}\sum_{i=1}^{N}(C_{Jk} - CJ_i)^2$$

(4)

**Results**

Assuming that the field is large enough (let's say over 10000 flowers) then normal distribution approximates well the sampling distribution of the number of colors for a given sample of flowers and any of the above mentioned methods can be applied to give the estimates of the desired parameter - number of the colors.

By applying Fisher's method, we need only the observed number of colors (C = 5, see Table 1) and the observed number of flowers (N = 44, see Table 1). Numerical solving of the Equation 1 gives α = 1.451889…; introducing of the values in the Equation 2 gives a variance of 0.938 for N = 44 and a variance of 1.007 for N → ∞.

Since the other methods are heavy computing ones, we will provide here only the results. Thus, Table 2 contains the estimates obtained with all applied methods.

Table 2. Estimates for the number of colors

| Method | Colors | Var(Colors) | Remarks |
|---|---|---|---|
| Fisher | 5 | 1.01 | small estimate; largest variance |
| Bootstrap | 4.695 | 0.213 | smallest estimate; smallest variance |
| Jackknife first order | 5.977 | 0.955 | large estimate; large variance |
| Jackknife second order | 5.9545 | 0.201 | large estimate; smallest variance |
| Jackknife third order | 5.932 | 0.270 | |

**Discussion**

It is expected that the Fisher's method to provide the largest variance. It uses smallest possible amount of information: only the number of flowers from the sample and the number of their colors. In order to provide a clear answer to the question from the title: "How many colors are actually in the field?" the inverse from the cumulative density function of the normal distribution at a 5% significance level should be calculated. Table 3 gives these estimates:

Table 3. How many different colors may have the flowers in the field?

| Method | Equation at 5% risk being in error | Solution | Answer |
|---|---|---|---|
| Fisher | $P\left(\frac{C-5}{1} \leq 1.96\right) = 0.975$ | $C \leq 6.98$ | $C_{max} \sim 7$ |
| Bootstrap | $P\left(\frac{C-4.695}{0.462} \leq 1.96\right) = 0.975$ | $C \leq 5.60$ | $C_{max} \sim 6$ |
| Bootstrap* | $P\left(\frac{C-4.695}{0.462} \leq 1.64\right) = 0.95$ | $C \leq 5.45$ | $C_{max} \sim 6$ |
| Jackknife first order | $P\left(\frac{C-5.977}{0.977} \leq 1.96\right) = 0.975$ | $C \leq 7.89$ | $C_{max} \sim 8$ |
| Jackknife second order | $P\left(\frac{C-5.9545}{0.6601} \leq 1.96\right) = 0.975$ | $C \leq 7.23$ | $C_{max} \sim 8$ |
| Jackknife third order | $P\left(\frac{C-5.932}{0.5194} \leq 1.96\right) = 0.975$ | $C \leq 6.95$ | $C_{max} \sim 7$ |
| * Computed according to (Efron & Tibshirani, 1993) | | | |

If the estimates are compared it can observed that the Bootstrap estimates are underestimates and first and second order Jackknife are overestimates (relatively to the supposition of negative binomial distribution - let's say the exact method).

There is no a general agreement about which method of numerical estimation should be used. The third order Jackknife proved the closest to the Fisher's method , but this is not a general conclusion. Thus, in (Severiano & others, 2011) authors obtained better results with Jackknife than with Bootstrap, and on the opposite, in (Smith & others, 1986) and (Hellmann & Fowler, 1999) the authors obtained better results with Bootstrap than with Jackknife.

The Fisher's method seems working very well in the investiated case - its result is on average of the others, and all results are very close to each other - and may be considered the golden test. Even more, the obtained results sustain the idea that the true distribution of the colors in the field is negative binomial. Let us note that the negative binomial distribution is often occurr in nature and it comes from a convolution of Poisson and Gamma distributions where the mixing distribution of the Poisson rate is the Gamma distribution (Jäntschi & others, 2011).

### Conclusions

The analysis for the number of colors of *Lycoris longituba* flowers shown that Fisher's method estimating the abundance agrees very well at its 95% confidence superior limit with the abundance estimated with third order Jackknife method. All investigated methods - Fisher's, Bootstrap, Jackknife of first, second and third orders provide estimates close to each other. This result sustains the Fisher's idea (adapted here from number of species to number of colors) that the observations follow the negative binomial distribution.

### Acknowledgments

The study was supported by POSDRU/89/1.5/S/62371 through a postdoctoral fellowship for L. Jäntschi.

The sponsors of the study had no role in the study design, data collection, data analysis, data interpretation or writing of this article.

Special thanks to the effective management of POSDRU/89/1.5/S/62371 for the visit study at 'Floriade' exhibition in Venlo, 2012 (L. Costoiu).